\documentclass{DISproc}
\pdfoutput=1
\usepackage{hepunits}
\usepackage{sidecap}

\begin{document}

\title{Gluon sivers and experimental considerations for TMDs}

\author{{\slshape Thomas Burton} \\ [1ex]
Brookhaven National Laboratory, Upton, NY 11973, USA}

% please enter the contribution ID for the DOI
\contribID{xy}

\doi  % if there is an online version we will register DOIs

\maketitle

\begin{abstract}

The study and characterisation of transverse-momentum-dependent distribution functions (TMDs) is a major goal of the Electron-Ion Collider (EIC) physics programme. The study of gluon TMDs poses a greater challenge than for quark TMDs in DIS measurements, as gluons do not directly couple to photons. The study of $D$ meson pairs has been proposed to provide access to gluon TMDs, but is demanding due to the rarity of $D$ production. Here, we discuss the feasibility of such a measurement, and touch upon wider issues to be considered when measuring TMDs at the EIC.

\end{abstract}

\section{Introduction}

The EIC is proposed as a next-generation collider facility, with unprecedented luminosity and the ability to study both nucleons and nuclei at a variety of energies.
There are two proposed realisations: eRHIC at Brookhaven National Lab (BNL), and ELIC at Jefferson Lab.
Building on the legacy of HERA, RHIC, CERN and others, it will deliver a broad programme of nucleon and nucleon structure studies with extremely high precision.
One of the key topics of interest at the EIC is the family of distribution functions known as TMDs.
TMDs add a transverse momentum component to conventional (one-dimensional) parton distribution functions (PDFs), providing a ``3D'' picture of partons in the the nucleon: TMDs essentially allow a tomographic imaging of the motion of partons at the femtoscale.
Work at HERMES and COMPASS, among others, has provided tantalising hints about these distributions.
However, data remains relatively scarce and a detailed understanding is far from achieved.
The high luminosity and broad kinematic reach of the EIC will provide the perfect environment for not just exploring these functions, but characterising them in exquisite detail.

The TMDs of quarks and gluons are of equal importance, but studying gluon distributions in electron-proton collisions is more challenging.
Unlike electrically charged quarks, gluons do not directly couple to photons, which mediate the e-p interaction.
A potential route to access gluon distributions is to study the production of pairs of $D$ mesons.
Here, we discuss the potential of such a method in measuring the gluon \emph{Sivers distribution}.
The Sivers distribution is a member of the TMD family that correlates the transverse momentum of partons in a nucleon with the transverse spin of the nucleon.
It has enjoyed a great deal of theoretical and experimental study in recent years, and as such stands as an exemplar for the wider TMD family.

\section{Studying the gluon Sivers function with $D$ meson pairs}

The methodology for studying the gluon Sivers function with $D$ mesons is detailed extensively in section 2.3 of \cite{Boer:2011fh}.
For certain kinematics, a charm-anticharm pair produced from a gluon (via the process $\gamma^{*}g\rightarrow q\bar{q}$) interacts like a single gluon.
Thus, the QCD interactions that give rise to the Sivers distribution in gluons will give rise to the same Sivers distribution in the charm quark pairs.
Quarks are not themselves detected, but their kinematics can be at least partly reconstructed from those of the $D$ mesons they produce.
As the Sivers distribution correlates spin and transverse momentum, a non-zero distribution results in a spin-dependent azimuthal asymmetry in the $D$ pair transverse momentum distribution around the virtual photon mediating the DIS interaction.
Measuring this asymmetry would allow the gluon Sivers distribution to be extracted.

This signature is exciting as it provides an avenue to measuring the Sivers function where few (or no) others are currently known to exist.
For example, dijet or dihadron production in proton-proton collisions, which would provide a natural environment for measuring gluon distributions due to the dominance of gluon interactions, are excluded due to the demonstration of factorisation breaking, precluding theoretical treatment.

While exciting, the measurement is not without its difficulties.
The rarity of $D$ meson production, coupled with the necessity to detect and reconstruct not one but two in a single event, makes it statistically challenging.
To investigate the feasibility of such a measurement, PYTHIA simulations were run to determine the event rates that could be expected at an EIC.

\section{PYTHIA simulation}

% Use \GeV not GeV so GeV is not italicised.
Charm events were generated with PYTHIA 6.416 for a high-energy EIC configuration: a \unit{20}{\GeVoverc} electron beam colliding with a \unit{250}{\GeVoverc} proton beam.
This configuration is most favourable for two reasons: an enhancement in the charm production cross section, and, at least in the BNL eRHIC design for the EIC, the maximum beam luminosity (see table \ref{tab:energies}).

\begin{wraptable}{l}{0.5\textwidth}
  \centering
  \begin{tabular}{c | c | c}
    \toprule
    Energy & eRHIC luminosity & $\sigma_{cc}$ \\
    \midrule
    5 x 100		&	$0.62\times\unit{\power{10}{33}}{\lumiunits}$	&	\unit{7.7}{\nanobarn} \\
    5 x 250		&	$9.70\times\unit{\power{10}{33}}{\lumiunits}$	&	\unit{13.3}{\nanobarn} \\
    20 x 250	&	$9.70\times\unit{\power{10}{33}}{\lumiunits}$	&	\unit{25.2}{\nanobarn} \\
    \bottomrule
  \end{tabular}
  \caption{eRHIC design luminosities and charm cross sections (from PYTHIA) for different eRHIC energy configurations (E$_{electron}$ x E$_{proton}$). Luminosity depends on the proton beam energy, but not the electron beam energy.}
  \label{tab:energies}
\end{wraptable}

The CT10 NLO PDF parameterisations \cite{Lai:2010vv} were used.
Radiative corrections were not explicitly included.
A cut on the DIS inelasticity of 0.01$<y<$ 0.95 was imposed and a single $Q^{2}$ bin of 1 to \unit{10}{\GeV^{2}} was used.
Events were required to contain at least two $D^{0}$ mesons.
The $D^{0}\rightarrow\pi$K decay channel was used, due to the large branching ratio (3.87\% \cite{pdg2010}) and the relative ease of experimentally reconstructing it.
Other $D^{0}$ decays are typically semileptonic or multi-hadon decays with three or more final-state products, significantly increasing the difficulty of, or excluding, reconstruction.
While other $D$ mesons ($D^{+}$, $D^{-}$, $D^{0*}$) could be used in an actual experiment, the increase in statistics from such decays was not sufficient to change the basic message of the Monte Carlo study: whether the analysis is feasible or not.
Therefore, the $D^{0}$$\rightarrow\pi K$ decay channel alone was used for simplicity.
$D^{0}$s generated directly from the DIS interaction and those generated via feed-down from heavier $D$ mesons, for example $D^{*+}$, were not distinguished.
Studies of the kinematics of a $D^{0}$ compared to that of the heavier $D$ generating it found little difference.
Thus, a feed-down $D^{0}$ is considered to provide as good a proxy for the charm quark kinematics as its parent.
The correlation between charm quark and $D$ meson kinematics was also examined.
They were typically found to be closely correlated in azimuth, by $\Delta\phi<$\unit{0.5}{\radian} in the virtual photon frame.

No detailed simulation of the effect of detector performance was done, but a simple acceptance cut, requiring all particles to be more than one degree from the beam direction, was imposed.

\section{Results and discussion}

\begin{wrapfigure}{r}{0.52\textwidth}
  \centering
  \includegraphics[width=0.5\textwidth]{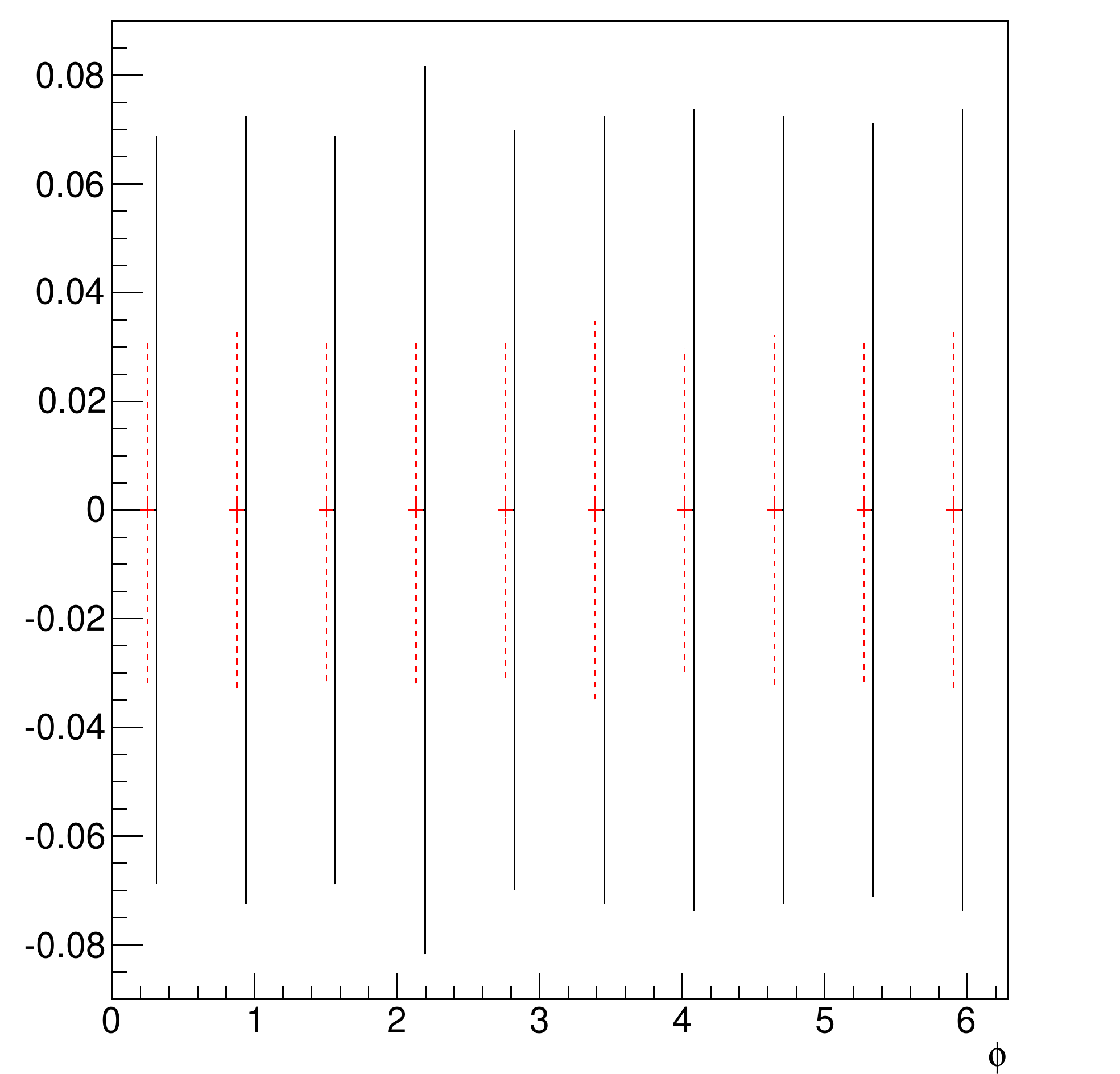}
  \caption{Estimated statistical uncertainties on the spin asymmetry in $D$ meson pair production with \unit{100}{\invfemtobarn} integrated luminosity for two bins in pair transverse momentum:  0.5 to \unit{1}{\GeVoverc} (red, dashed) and 1 to \unit{2}{\GeVoverc} (black, solid). Bin-to-bin variations are due to limited statistics in the Monte Carlo sample.}
  \label{fig:errors}
\end{wrapfigure}

$D^{0}$ pairs were binned simultaneously in azimuthal angle around the virtual photon direction and transverse momentum, and the statistical uncertainty in each bin estimated as $\sqrt{N}$ (for $N$ counts).
Uncertainties were rescaled to represent \unit{100}{\invfemtobarn} of data.
As the measurement is a single-spin asymmetry, error bars were also divided by proton beam polarisation, which was assumed to be 70\% (the eRHIC design polarisation).
The resulting statistical uncertainties are shown in Figure \ref{fig:errors} for two bins in $D$ meson pair transverse momentum and 10 bins in azimuthal angle.

As seen in the figure, uncertainties of a few to several percent can be anticipated, depending on the pair transverse momentum range.
The potential asymmetries derived from the model in \cite{Boer:2011fh} depend on the the pair transverse momentum and can exceed 20\% for $k_{perp} > $\unit{1}{\GeVoverc}.
This indicates that measuring $D$ pair production is indeed feasible as an avenue to measuring the gluon Sivers function.
At its maximum luminosity, and conservatively assuming an operational efficiency of 50\%, eRHIC would accumulate about \unit{3}{\invfemtobarn} per week.
\unit{100}{\invfemtobarn} would therefore require close to entire year of running at eRHIC.
The eRHIC physics programme would generally demand multiple energy configurations to be run in each year, so such an amount would likely represent a multi-year goal.

\begin{figure}
  \centering
  \includegraphics[width=0.5\textwidth]{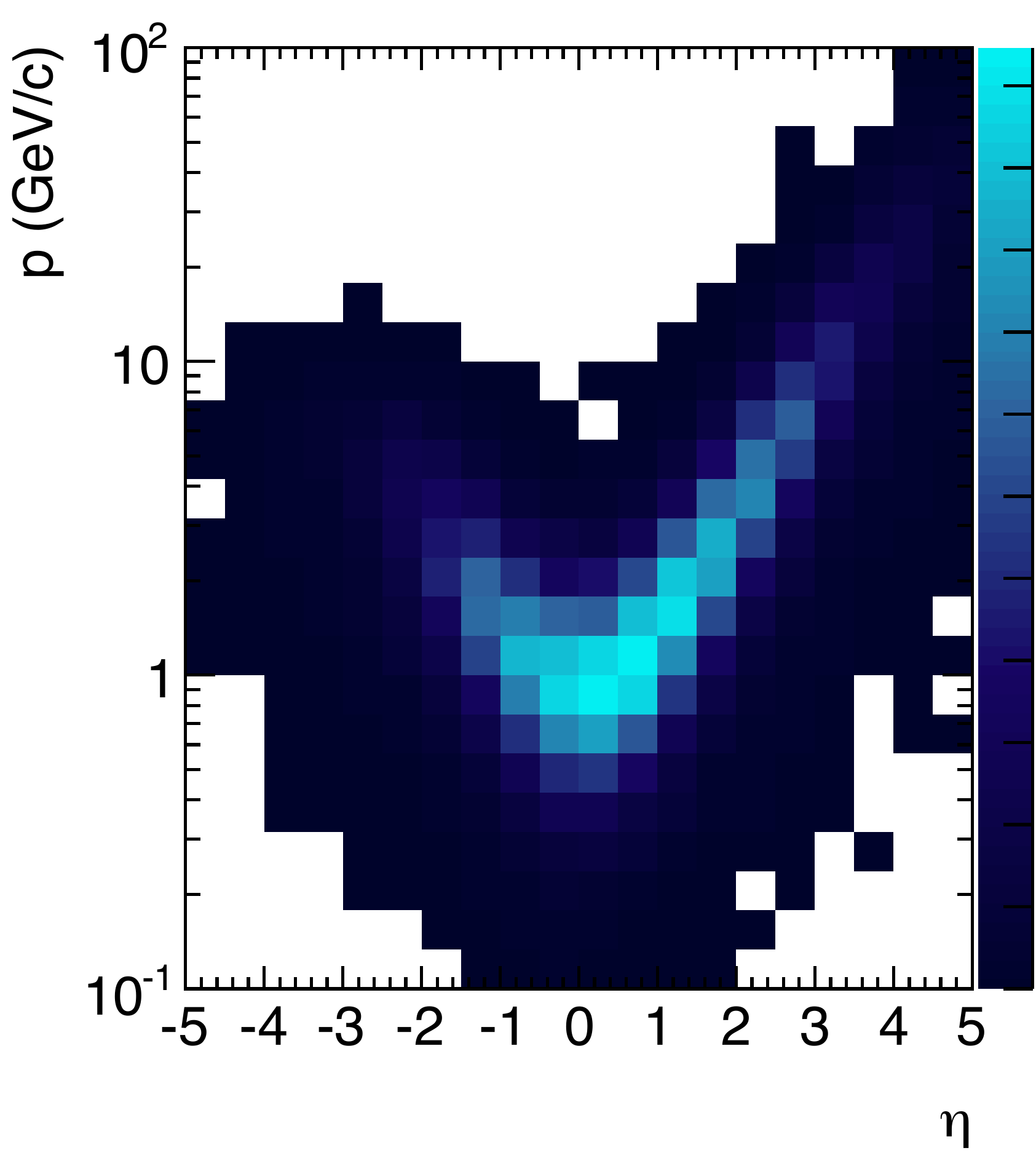}
  \caption{Distriubution of $\pi$ from $D^{0}$ decays in momentum and pseudorapidity (colour online).}
  \label{fig:pions}
\end{figure}

The measurement of $D$ mesons highlights key performance requirements for an EIC detector that apply more generally to other semi-inclusive and exclusive measurements at an EIC.
Perhaps most important is the requirement for precise tracking and particle identification (PID) over a wide range in rapidity.
As is seen in Figure \ref{fig:pions}, not only are hadrons produced in the proton beam direction, but also at significant rates in the central region and the electron beam direction.
Hadron momenta are typically modest in the central region (less than a few \GeVoverc), but extend to to several tens of {\GeVoverc} in the proton beam direction.
Therefore PID detectors capable of distinguishing hadrons up to high momentum will be required in this region.
Furthermore, in common with inclusive events, the scattered electron should be detected and measured precisely to determine the DIS kinematics.
An EIC detector should therefore provide high momentum and angular resolution and hadron-electron separation over a several units of rapidity.
Whilst doing so, the material budget of the tracking detectors must be kept low enough to allow measurement of electrons down to low momenta (less than \unit{1}{\GeVoverc}), where multiple scattering becomes an issue.

\section{Bibliography}

\bibliography{burton_thomas_gluon_sivers.bib}

\providecommand{\href}[2]{#2}\begingroup\raggedright\begin{thebibliography}{1}

\bibitem{Boer:2011fh}
D.~Boer, M.~Diehl, R.~Milner, R.~Venugopalan, W.~Vogelsang, {\em et~al.}
\newblock
\href{http://arxiv.org/abs/1108.1713}{{\ttfamily arXiv:1108.1713 [nucl-th]}}.
%%CITATION = ARXIV:1108.1713;%%.

\bibitem{Lai:2010vv}
H.-L. Lai, M.~Guzzi, J.~Huston, Z.~Li, P.~M. Nadolsky, {\em et~al.}
\newblock \href{http://dx.doi.org/10.1103/PhysRevD.82.074024}{Phys.Rev.
  {\bfseries D82} (2010) 074024},
\href{http://arxiv.org/abs/1007.2241}{{\ttfamily arXiv:1007.2241 [hep-ph]}}.
%%CITATION = ARXIV:1007.2241;%%.

\bibitem{pdg2010}
K.~Nakamura and P.~D. Group.
\newblock Journal of Physics G: Nuclear and Particle Physics {\bfseries 37}
  no.~7A, (2010) 075021.
  \url{http://stacks.iop.org/0954-3899/37/i=7A/a=075021}.

\end{thebibliography}\endgroup

\end{document}